\documentclass[epj]{svjour}
\usepackage[11pt]{extsizes}
\usepackage{graphics}

\usepackage[paper=a4paper,margin=1.0in]{geometry}

\pdfoutput=1 
\usepackage{hyperref}
\usepackage{geometry}
\usepackage{mathtools}
\usepackage{graphicx}
\usepackage{latexsym}

\usepackage{amssymb}
\usepackage{amscd,amssymb}
\usepackage[arrow,matrix]{xy}
\usepackage{graphicx}
\usepackage{epstopdf}
\usepackage{setspace,cite}
\usepackage{amscd,amsmath,amssymb}

\newcommand{\sech}{\mbox{sech}}

\newcommand{\bea}{\begin{eqnarray}}
\newcommand{\eea}{\end{eqnarray}}
\newcommand{\bes}{\begin{subequations}}
	\newcommand{\ees}{\end{subequations}}

\usepackage{lipsum}
\usepackage{fancyhdr}
\begin{document}
\title{Benjamin-Ono equation: Rogue waves, generalized breathers, soliton bending, fission, and fusion}
\titlerunning{Benjamin-Ono equation: Rogue waves, generalized breathers, and solitons}

\author{\bf Sudhir Singh\inst{1} \and K. Sakkaravarthi\inst{2,3}$^*$ \and K. Murugesan\inst{1} \and R. Sakthivel\inst{4}$^*$ 
\thanks{Email: sudhirew@gmail.com, ksakkaravarthi@gmail.com, murugu@nitt.edu, krsakthivel@yahoo.com \newline $^*$Corresponding authors}}

\authorrunning{Sudhir Singh {\it et al}} 

\institute{Department of Mathematics, National Institute of Technology, Tiruchirappalli -- 620015, Tamil Nadu, India. 
	\and
	Department of Physics, National Institute of Technology, Tiruchirappalli -- 620015, Tamil Nadu, India 
	\and
	Centre for Nonlinear Dynamics, School of Physics, Bharathidasan University, Tiruchirappalli -- 620024, Tamil Nadu, India. 
	\and
	Department of Applied Mathematics, Bharathiar University, Coimbatore -- 641046, Tamil Nadu, India. 
}
%
%
\date{{\bf Published Journal Reference: \href{https://doi.org/10.1140/epjp/s13360-020-00808-8}{\bf Eur. Phys. J. Plus 135 (2020) 823.}}}
%
\abstract{
In this work, we construct various interesting localized wave structures of the Benjamin-Ono equation describing the dynamics of deep water waves. Particularly, we extract the rogue waves and generalized breather solutions with the aid of bilinear form and by applying two appropriate test functions. Our analysis reveals the control mechanism of the rogue waves with arbitrary parameters to obtain both bright and dark type first and second-order rogue waves. Additionally, a generalization of the homoclinic breather method, also known as the three-wave method, is used for extracting the generalized breathers along with bright, dark, anti-dark, rational  solitons. Interestingly, we have observed the manipulation of breathers along with soliton interaction, bending, fission, and fusion. Our results are discussed categorically with the aid of clear graphical demonstrations. 
} 
\maketitle
\setstretch{1.12}

\section{Introduction}
The study of water waves gains tremendous interest over more than two centuries; theoretical and experimental investigations are continuously increasing to understand the dynamics of these waves \cite{ML-book}. These wave equations are modeled using prototype ordinary/partial/delay/fractional differential equations \cite{rs1,Yang-book}. The nonlinear dynamics of waves associated with such model equations explore several exciting phenomena including wave-mixing/breaking and interaction of several localized structures like solitons, breathers, lump solutions, and rogue waves, which have several applications in fluid dynamics, plasma, Bose-Einstein condensate, fiber optics and even in finance \cite{Yang-book,bgo}. 
These wave models also studied in the sense of weak type fractional stochastic equations, where the analysis of these models are more general \cite{rs2}. One of the most agreeable localized structure is soliton. It can be viewed as a classical solution structures of the integrable models. It is also well-known that the stability or identity-preserving nature of these solution structures even after the collisions enable them to have phenomenal applications in diverse areas of science and technology. Notably, these nonlinear wave properties help tackle the behaviour of real wave structures, including DNA, plasma, wave transmissions in the optical fiber, and many more. 

One natural question arises: Why do  other localized solution structures apart from solitons, such as rogue waves and breathers, multi-shock waves, and lumps, are required? It can be explained as the presence of several other highly unstable practical phenomena in nature; in those situations, these nonlinear wave structures are necessary to understand them thoroughly. Rogue waves are relatively a new kind of localized structures, and they are also known as ``monster waves, killer waves, extreme waves" and ``freak waves" \cite{rogue-book}. Their behaviour is mysterious, and it can also be associated with the chaotic phenomenon. A famous saying for rogue waves is `coming from nowhere and disappear with no trace' \cite{nak}, because of the reason that the rogue waves are temporally and spatially localized disturbance and amplitude is increasing on the background by a few orders of magnitude \cite {nak, PRX-rogue}, which can also be considered as a possible explanation for its chaotic behavior. Further, these rogue waves appear for several reasons, including a universal route of modulation instability and wave synchronization \cite{PRX-rogue}. These waves appear as substantial, large localized structures compared to other localized waves. Their height is approximately (more than twice the significant wave height) two or more times the height of the surrounding waves. In the ocean, their occurrence damages ships and oil drilling platforms. However, the appearance of rogue waves is not limited to the sea but in finance \cite {zyn}, plasma \cite {wmm}, superfluid \cite {vbe}, Bose-Einstein condensate \cite {yvb} and well known optical rogue waves \cite {drs}. A rigorous treatment for rogue waves, including modeling and experimental observations, is being done continuously with different models \cite{bgu, ckh, PRX-rogue,AkhJOpt}. The exact solutions of the nonlinear evolution equations help significantly understand the dynamics of the waves and these solutions with different physical structures have phenomenal applications in a broader range of science and engineering \cite{rs3,rs4}. 

Being motivated by the increasing interest in the rogue waves, we devote our investigation in understanding them in a well-known Benjamin-Ono equation. Before moving on to the main study, we briefly revisit the rogue wave solution of the following standard nonlinear Schr\"odinger (NLS) equation \cite{Kiv-book}: 
\begin{equation} \label{eq:1}
i u_t+u_{xx}+2 | u|^2  u=0,
\end{equation}
The above NLS model appears as a governing equation in different fields ranging from nonlinear optics, Bose-Einstein condensates, plasma physics, etc. For the above focusing type NLS equation (\ref{eq:1}), the Peregrine breather (simplest rational solution) is available from the literature \cite{aca} as
\begin{equation}
u(x,t)=\alpha\left({\dfrac{4(1+ 4 i \alpha ^2 t)}{1 + 4 \alpha ^2 x^2 + 16 \alpha ^4 t^2}}-1\right) \exp(2i \alpha ^2 t),
\end{equation}
The above mentioned doubly-localized Peregrine breather solution (2) is graphically demonstrated in Fig. 1. D. H. Peregrine first proposed this kind of solution for the focusing Schr\"odinger equation \cite{dhp}. This solution is localized in space as well as in time and rises over a uniform background of Stokes wave solution $u_s= \exp(i2 \alpha ^2 t)$. Later various rational solutions for different models were constructed using different analytical methods, including the famous inverse scattering transform, Darboux transformation, Hirota method, dressing method, and several ansatz approaches \cite{rogue-book}. 
\begin{figure}[h] 
	\begin{center} 
		\includegraphics[width=0.81\linewidth]{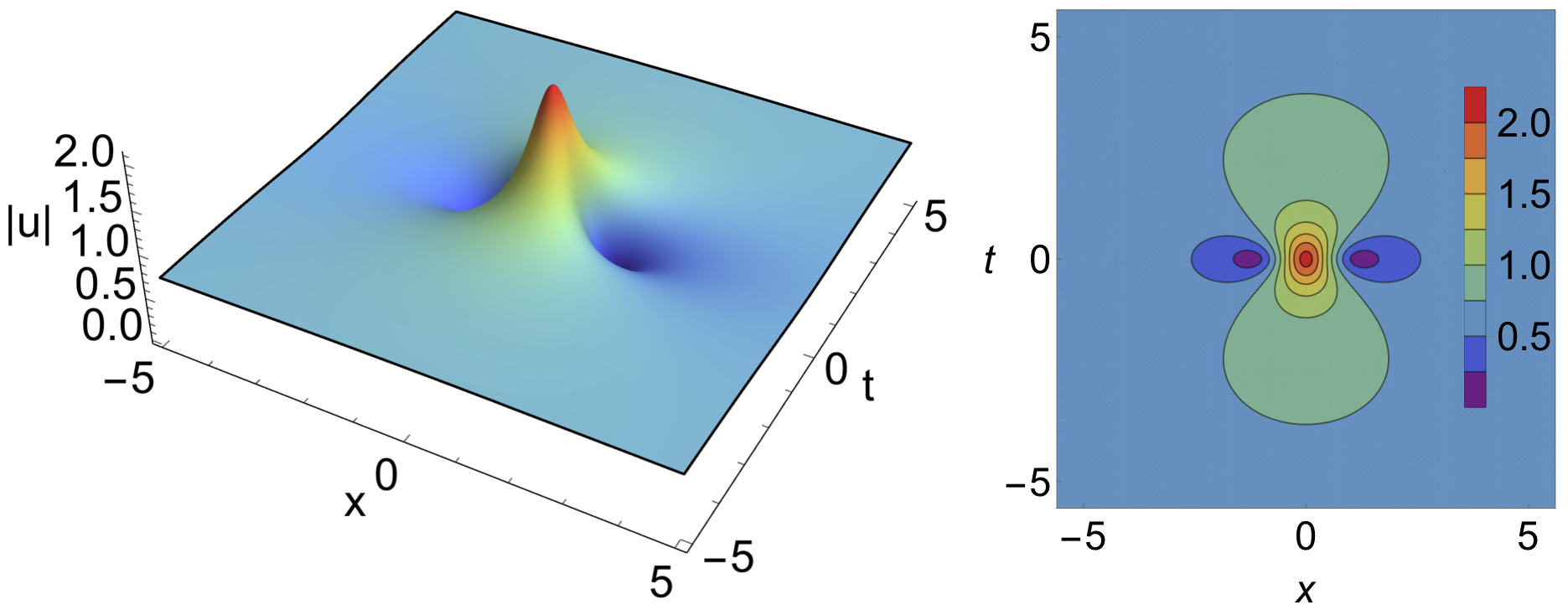}
		\caption{Intensity and contour plot of the Peregrine breather solution (2) for $\alpha=0.7$.}
	\end{center} \label{fig-pere}
\end{figure}

Now, we consider the following form of the Benjamin-Ono (BO) equation:
\begin{equation}
u_{tt} + 2\beta u_{x}^2+ 2\beta u u_{xx}  + \gamma u_{xxxx} =0, \label{eq1.1}
\end{equation}
where $\beta$ and $\gamma$ are arbitrary nonlinearity and dispersion coefficients, respectively. The Benjamin-Ono equation describes one-dimensional internal waves in deep water \cite{tbb,hon}. It is mathematically a famous nonlinear partial integrodifferential equation possessing integrability and solvable by inverse scattering transform as well as possess B\"acklund transformation, Painlev\'e singularity analysis and also admitting $N$-soliton solutions \cite{asf,kmc,kmt,rij} and have infinite conserved quantities apart from special rational and periodic solutions \cite{ana,jsat}. Its various other solutions, including nonlocal symmetries \cite{yli} and rogue wave solutions \cite{wli}, are available in the literature. Our aim of the present work is to explore various physically significant solutions such as rogue waves, breathers, and solitons with a newly developed method and study their control mechanism along with the interaction, bending, fusion, and fission of nonlinear waves especially solitons. 

Now by using a bilinearizing transformation, we write the BO equation into a compact bilinear form and then solve it by introducing a polynomial test function of the required order. For this purpose, by making use of the following transformation 
\bea u(x,t)= u_0 + \dfrac{6 \gamma}{\beta}(ln \, f)_{xx}, \label{trans}\eea 
where $u_0$ is an arbitrary background while $f(x,t)$ is a arbitrary function to be determined, the BO equation (\ref{eq1.1}) is transformed into the following Bilinear form:
\begin{equation}
(D_t^2 + 2 u_0 \beta D_x^2 + \gamma D_x ^4) f \cdot f=0. \label{eq2.1}
\end{equation} 
Here $D$ represents the standard Hirota differential operator \cite{wli,NLD2017BO,Hirota-book} and it can be defined as 
$$D_x ^ a D_y^b f \cdot g = \left ( \dfrac{\partial }{\partial x} - \dfrac{\partial }{\partial x'} \right ) ^ a \left ( \dfrac{\partial }{\partial y} - \dfrac{\partial }{\partial y'} \right ) ^ b  f(x,y,t) \cdot g(x',y',t')  | _{(x,y,t)=(x',y',t')}.$$
In recent years, localized structures such as rogue waves, breathers, and lump solutions arising in various nonlinear models attract more concentration and they become interesting both in mathematical and physical perspective due to their occurrence in diversified areas like plasmas, optics, Bose-Einstein condensate, and financial systems \cite{rogue-book,nak, PRX-rogue}.\\

The present work is organized as follows. In Sec. 2, the first and second-order rogue wave solutions are constructed using the Hirota bilinear form and polynomial test functions \cite{ylm}. In Sec. 3, the generalized breather solutions of different wave structures are obtained by using the three-wave method \cite{wta2} along with their evolution dynamics. And conclusions are provided in the final section.

\section{Rogue Waves}
In this section, we construct the first and second-order rogue wave solutions using the bilinear form \eqref{eq2.1} and appropriate polynomial test functions and investigate their evolution with respect to different control parameters. 

\subsection{Rogue wave solution of order-one}
To extract the rogue wave solution of order-one, we choose the form of $f(x,t)$ as \cite{ylm}:
\begin{equation}
f(x,t) = k_0 + ( \alpha_1 x + \beta _1 t) ^2 + ( \alpha _2 x + \beta _2 t) ^2. \label{eq1}
\end{equation}
Substituting the above form of $f$  (\ref{eq1}) into bilinear equation (\ref{eq2.1}) and collecting the coefficients of different powers of $\{ x^i y^j,~ i, j=0,1,2 \}$, we get the following set of equations:\\
\begin{subequations}
	\begin{align}
	\text{ Coefficient of } x^2 & : ( -4 u_0 \alpha _1 ^4 \beta - 8 u_0 \alpha_1 ^2 \alpha _ 2 ^2 \beta - 4 u_0 \alpha _2 ^4 \beta - 2 \alpha _1 ^2 + \beta _ 1 ^ 2 + 2 \alpha_2 ^2 \beta _1 ^2 - 8 \alpha_1 \alpha_2 \beta _1 \beta _2 \notag  \\
	&  + 2 \alpha _1 ^2 \beta _2 ^2 - 2 \alpha_2 ^2 \beta _2 ^2 ) =0, \label{eq2a} \\
	\text{ Coefficient of } t^2 & : ( - 4 u_0 \alpha_1 ^2 \beta \beta _1 ^2 + 4 u_0 \alpha _2 ^2 \beta \beta _1 - 2 \beta _1 ^4 - 16 u_0 \alpha _1 \alpha_2 \beta \beta _1 \beta _2 + 4 u_0 \alpha_1 ^2 \beta \beta _2 ^2 \notag  \\ 
	&  - 4 u_0 \alpha_2 ^2 \beta \beta _2 ^2 - 4 \beta _1 ^2  \beta _2 ^2 -2 \beta _2 ^4 ) =0, \label{eq2b} \\
	\text{ Coefficient of } t x & : (- 8 u_0 \alpha _1 ^3 \beta \beta _1 - 8 u_0 \alpha_1 \alpha_2 ^2 \beta \beta _1 - 4 \alpha _1 \beta _1 ^3 - 8 u_0 \alpha_1 ^2 \alpha_2 \beta \beta _2 - 8 u_0 \alpha_2 ^3 \beta \beta _2 \notag \\
	&  - 4 \alpha_2 \beta _1 ^2 \beta _2 - 4 \alpha_1 \beta _1 \beta _2 ^2 - 4 \alpha_2 \beta _2 ^3 ) =0, \label{eq2c} \\
	\text{ Constants } &  : (4 k_0 u_0 \alpha_1 ^2 \beta + 4 k_0 u_0 \alpha_2 ^2 \beta + 2 k_0 \beta _1 ^2 + 2 k_0 \beta _2 ^2 + 12 \alpha_1 ^4 \gamma + 24 \alpha_1 ^2 \alpha_2 ^2 \gamma + 12 \alpha_2 ^4 \gamma ) =0. \label{eq2d}
	\end{align}
\end{subequations}
Solving the above system of equations (\ref{eq2a})-(\ref{eq2d}), we obtain the following relations among the parameters resulting to the first-order rogue wave solution as 
\begin{align}
u_0 \beta >0, \quad 	\beta _1 & = \pm \sqrt{2 u_0 \beta } \alpha_2 , \quad \beta _2 = \mp \sqrt{2 u_0 \beta } \alpha_1 , \quad k_0 = \dfrac{-3 (\alpha_1 ^2 +\alpha_2) ^2 \gamma}{2 u_0 \beta }. \label{eq3}
\end{align}
From Eqs. (\ref{eq3}) and (\ref{eq1}), we get the explicit form of $f$ as 
\begin{equation}
f(x,t) = \left( \alpha_1 x + \sqrt{2 u_0 \beta } \,  \alpha_2 t\right) ^2 + \left(\alpha_2 x - \sqrt{2 u_0 \beta } \, \alpha_1 t \right) ^2 - \dfrac{3 ( \alpha_1 ^2 + \alpha_2 ^2 ) \gamma }{2 u_0 \beta}. \label{eq4} 
\end{equation}
Thus, we can obtain the  first-order rouge wave solution from \eqref{trans} and \eqref{eq4} as 
\begin{equation}
u(x,t)= u_ 0 + \dfrac{24 u_0\gamma ( -2 u_0 x ^2 \beta + 4 t^2 u_0 ^2 \beta ^2 - 3 \gamma)  }{(2 u_0 x ^2 \beta + 4 t^2 u_0 ^2 \beta ^2 - 3 \gamma )^2 } ,\label{eq5}
\end{equation}
under the constraint condition $u_0 \beta >0$. The above solution carries three arbitrary parameters $u_0,~\beta$, and $\gamma$. It is of natural interest to understand importance and roles of these arbitrary parameters in defining the dynamics of the rogue wave solution \eqref{eq5}. We can obtain two types of rogue waves, namely bright and dark, by tuning these parameters as shown in Fig. 2 by retaining the necessary condition $u_0 \beta >0$. We mainly obtain bright single peak doubly localized excitation with the choice $u_0=-1.05$, $\beta=-1.01$, and $\gamma=1.05$. On the other hand, a dark type first-order rogue wave structure is depicted for another choice $u_0=1.05$, $\beta=0.3$, and $\gamma=-0.05$. 
\begin{figure}[h] 
	\begin{center} 
		\includegraphics[width=0.891\linewidth]{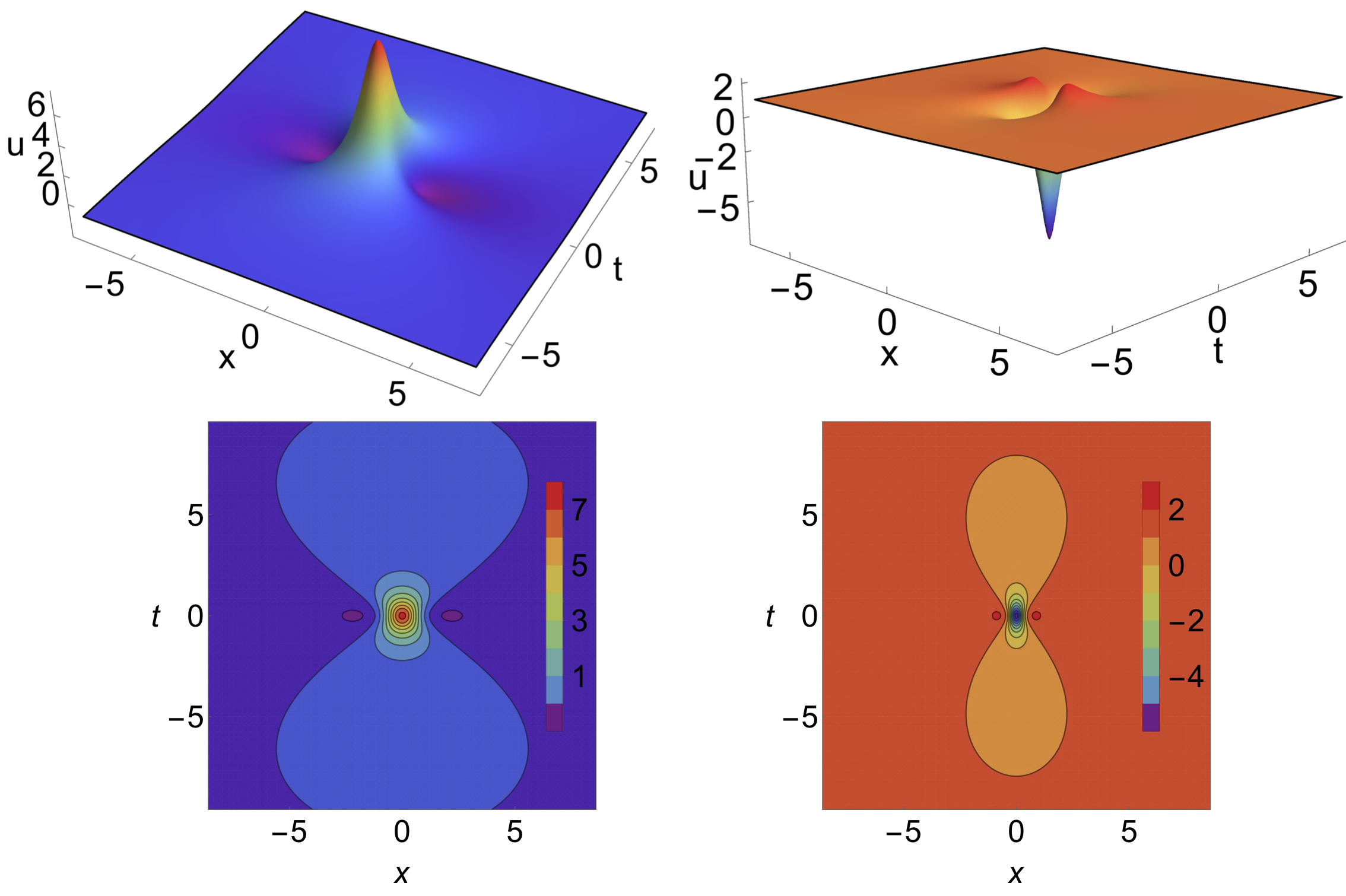}
		\caption{Bright (left panel) and dark (right panel) type first-order rogue waves through solution \eqref{eq5}. The bottom panel shows the corresponding contour plots.}
	\end{center} \label{fig-rogue1}
\end{figure}

For a much clear inference of the arbitrary parameters, we have demonstrated their role graphically for the bright rogue wave with three different sets of values in Fig. 3. Our analysis shows that the increase in the magnitude of $\beta$ decreases the width of the rogue waves, while the $\gamma$ parameter is altering its width in direct proportion with an appreciable change in their tail and without affecting the amplitude. However, the parameter $u_0$ is much simpler, increasing the amplitude of the rogue wave and a shift from its constant background. Importantly, there occurs an appreciable change in the significant wave height (amplitude) of the amplitude. Similar effects can also be observed in the dark rogue wave case, which is shown in Fig. 4, where the depth/darkness, background, width, and tail of the dark rogue waves are controlled by tuning $\beta$, $\gamma$, and $u_0$ parameters.
\begin{figure}[h] 
	\begin{center} 
		\includegraphics[width=1.05\linewidth]{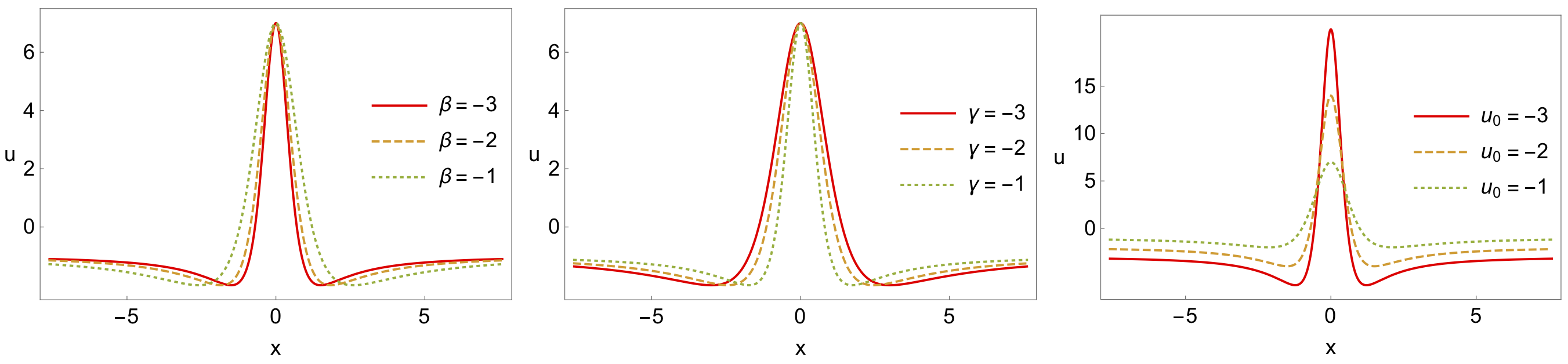}
		\caption{Impact of $\beta$, $\gamma$ and $u_0$ parameters in the first-order bright rogue wave \eqref{eq5} for fixed values of other parameters. }
	\end{center} \label{fig-rogue1a}
\end{figure}
\begin{figure}[h] 
	\begin{center} 
		\includegraphics[width=1.05\linewidth]{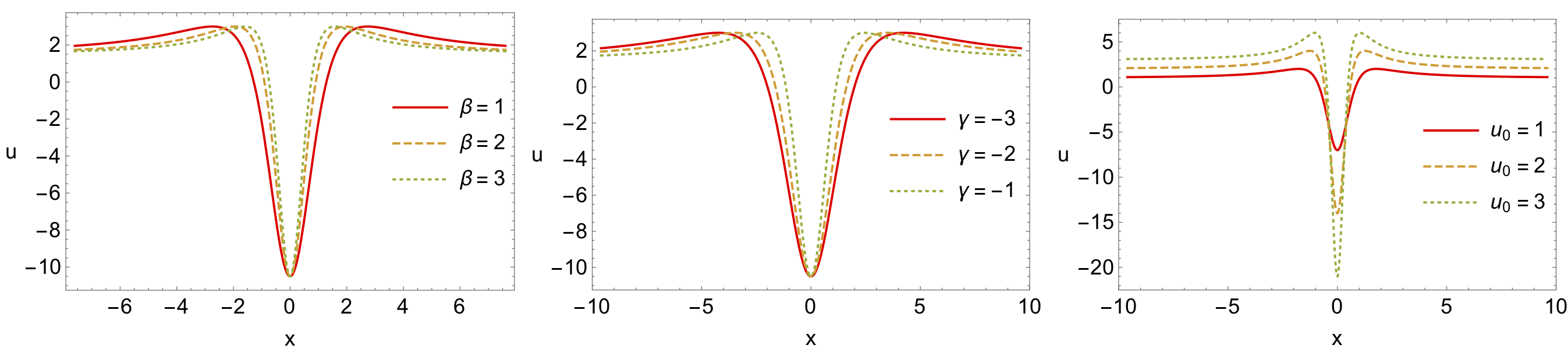}
		\caption{Impact of $\beta$, $\gamma$ and $u_0$ parameters in the first-order dark rogue wave \eqref{eq5} for fixed values of other parameters. }
	\end{center} \label{fig-rogue1b}
\end{figure}

\subsection{Rogue wave solution of order-two}
The next natural step is to construct and look for higher-order rogue waves and their dynamical behaviour in the present BO equation. Here we construct a simplest higher-order rogue wave, which is the rogue wave of order two. For this purpose, we have  adopted the following polynomial test function \cite{ylm}: 
\begin{equation}
f(x,t) = \beta_1 + \left( \beta _2 x ^2 + \beta _3 t^2 \right) ^3 + \beta_ 4 x ^4 + \beta _5 x ^2 t^2 + \beta _6 t^4 + \beta _7 x ^2 + \beta _8 t^2, \label{eq6}
\end{equation}
where $\beta_j (j=1,2,\ldots, 8)$ are parameters of second-order rogue wave solution, will be determined later. Substituting (\ref{eq6}) into the bilinear form (\ref{eq2.1}) and equating to zero the coefficients of $\{ x^m y^n, \quad m,n= 0,1,2,3,4,5,6,7,8\} $, we obtained the following  algebraic nonlinear system of equations:
\begin{subequations}
	\bea 
	\text{Coefficient of } x ^{10} :& -12 u_0 \beta \beta _2 ^6 + 6 \beta _2 ^5 \beta _3 =0, \label{eq7a}\\
	\text{Coefficient of } t^2 x ^{8} :& -36 u_0 \beta \beta _2 ^5 \beta _3 + 18 \beta _2 ^4 \beta _3 ^2 =0, \label{eq7b}\\
	\text{Coefficient of } t^4 x ^{6} :& -24 u_0 \beta \beta _2 ^4 \beta _3 ^2 +12 \beta _2 ^3 + \beta _3 ^3=0, \label{eq7c}\\
	\text{Coefficient of } t^6 x ^{4} :&
	24 u_0 \beta \beta _2 ^3 \beta _3^3 - 12 \beta _2 ^2 \beta _3 ^4 =0, \label{eq7d}\\
	\text{Coefficient of } t^8 x ^{2} :& 36 u_0 \beta \beta _2 ^2 \beta _3 ^4 - 18 \beta_2 \beta _3 ^5 =0, \label{eq7e} \\
	\text{Coefficient of } t^{10} :& 12 u_0 \beta \beta _2 \beta _3 ^5 - 6 \beta _3 ^6=0, \label{eq7f} \\
	\text{Coefficient of }  x ^{8} :& -12u_0 \beta \beta _2 ^3 \beta _4 + 6 \beta _2 ^2 \beta _3 \beta _4 + 2 \beta _2 ^3 \beta _5 + 180 \beta _2 ^6 \gamma =0, \label{eq7g}\\
	\text{Coefficient of } t^2 x ^{6} :&  -48 u_0 \beta \beta2 ^2 \beta _3 \beta _4 + 36 \beta _2 \beta _3 ^2 \beta _4 + 16 u_0 \beta \beta _2 ^2 \beta _5 -12 \beta _2 ^2 \beta _3 \beta_5 + 12 \beta _2 ^3 \beta_6 \notag \\
	&  + 144 \beta _2 ^5 \beta _3 \gamma =0, \label{eq7h} \\
	\text{Coefficient of } t^4 x ^{4} :& 
	-12 u_0 \beta \beta _2 \beta _3 ^2 \beta _4 + 30 \beta _3 ^3 \beta _4  - 12 u_0 \beta  \beta _2 ^2 \beta _3 \beta_5 - 6 \beta_2 \beta _3 ^2 \beta _5 + 60 u_0 \beta \beta _2 ^3 \beta _6 - 6 \beta _2 ^2 \beta _3 \beta _6\notag \\
	&  - 72 \beta _2 ^4 \beta _3 ^2 \gamma =0, \label{eq7i}\\
	\text{Coefficient of } t^6 x ^{2} :&  24 u_0 \beta \beta _3 ^3 \beta _4 - 24 u_0 \beta \beta _2 \beta _3 ^2 \beta _5 + 8 \beta _3 ^3 \beta _5 + 72 u_0\beta \beta _2 ^2 \beta _3 \beta _6 - 24 \beta _2 \beta _3^2 \beta _6 + 144 \beta _2 ^3 \gamma =0, \label{eq7j} \\
	\text{Coefficient of } t^8  :&  4 u_0 \beta \beta _3 ^3 \beta _5 + 12 u_0 \beta \beta _2 \beta _3 ^2 \beta_6 - 6 \beta _3 ^3 \beta _6 + 180 \beta _2 ^2 \beta _3 ^4 \gamma =0, \label{eq7k}\\
	\text{Coefficient of } x ^{6}: & -8 u_0 \beta \beta _4 ^2 + 2 \beta _4 \beta _5 + 16 u_0 \beta \beta _2 ^3 \beta_7 + 6 \beta _2 ^2 \beta _3 \beta _7 + 2 \beta _2 ^3 \beta _8 + 48 \beta _2 ^3 \beta _4 \gamma =0, \label{eq7l} \\
	\text{Coefficient of } t^2 x ^{4} :&  -4 u_0  \beta \beta _4 \beta _5 - 2 \beta _5 ^2 + 12 \beta _4 \beta _6 - 12 u_0 \beta \beta _2 ^2 \beta _3 \beta _7 + 36 \beta _2 \beta _3 ^2 \beta _7 + 60 u_0 \beta \beta_2 ^3 \beta _8 \notag \\
	&  - 12 \beta _2 ^2 \beta _3 \beta _8 + 432 \beta _2 ^2 \beta _3 \beta _4 \gamma - 240 \beta _2 ^3 \beta _5 \gamma =0, \label{eq7m} \eea \bea 
	\text{Coefficient of } t^4 x ^{2} :&  -4 u_0 \beta beta _5 ^2 + 24 u_0 \beta \beta _4 \beta _6 - 2 \beta _5 \beta _6 - 24 u_0 \beta \beta _2 \beta _3^2 \beta _7 + 30 \beta _3 ^3 \beta _7 + 72 u_0 \beta \beta _2 ^2 \beta _3 \beta _8\notag \\
	& - 6 \beta _2 \beta _3 ^2 \beta _8 - 72 \beta _2 \beta _3 ^2 \beta _4 \gamma - 72 \beta _2 ^2 \beta 3 \beta _5 \gamma + 360 \beta _2 ^2 \beta _6 \gamma =0, \label{eq7n}\\
	\text{Coefficient of } t ^{6} :&  4 u_0 \beta \beta _5 \beta _6 - 4 \beta _6 ^2 + 4 u_0 \beta \beta _3 ^3 \beta _7 + 12 u_0 \beta \beta _2 \beta _3^2 \beta _8 + 8 \beta _3 ^3 \beta _8 +24 \beta_3^3 \beta_4 \gamma+72 \beta_2 \beta_3^2 \beta _5 \gamma \notag \\
	&+72\beta_2^2 \beta _3 \beta _6 \gamma=0, \label{eq7o}\\
	\text{Coefficient of } x ^{4} :&  60 u_0 \beta \beta _1 \beta _2 ^3 + 6 \beta _1 \beta _2 ^2 \beta _3 - 4 u_0 \beta \beta _4 \beta _7 + 2 \beta _5 \beta _7 + 2 \beta _4 \beta _8 + 72 \beta _4 ^2 \gamma - 240 \beta _2 ^3 \beta _7 \gamma =0, \label{eq7p}\\
	\text{Coefficient of } x ^{2} :&  24 u_0 \beta \beta _1 \beta _4 + 2 \beta _1 \beta _5 - 4 u_0 \beta \beta _7 ^2 + 2 \beta _7 \beta _8 + 360 \beta _1 \beta _2 ^3 \gamma - 24 \beta _4 \beta _7 \gamma = 0, \label{eq7q} \\
	\text{Coefficient of } t^2  x ^{2} :& 72 u_0  \beta \beta_1 \beta _2 ^2 \beta _3 + 36 \beta _1 \beta _2 \beta _3 ^2 - 8 u_0 \beta \beta _5 \beta _7 + 12 \beta _6 \beta _7 + 24 u_0 \beta \beta _4 \beta _8 - 4 \beta _5 \beta _8 - 24 \beta _4 \beta _5 \gamma \notag \\
	&  - 72 \beta _2 ^2 \beta _3 \beta _7 \gamma + 360 \beta_2 ^3 \beta _8 \gamma =0,  \label{eq7r}\\
	\text{Coefficient of } t ^{4} :&  12 u_0 \beta \beta_1 \beta _2 \beta _3^2 + 30 \beta _1 \beta _3 ^3 + 4 u_0 \beta \beta _6  \beta _7 + 4 u_0 \beta \beta _5 \beta _8 - 2 \beta _6 \beta _8 + 12 \beta _5 ^2 \gamma + 24 \beta _4 \beta _6 \gamma \notag \\
	&  + 72 \beta _2 \beta _3 ^2 \beta _7 \gamma + 72 \beta _2 ^2 \beta _3 \beta _8 \gamma =0, \label{eq7s} \\
	\text{Coefficient of } t ^{2} :&  4 u_0 \beta \beta _1 \beta _5 + 12 \beta _1 \beta _6 + 4 u_0 \beta \beta _7 \beta_ 8 - 2 \beta _8 ^2 + 72 \beta _1 \beta _2 ^2 \beta _3 \gamma + 24 \beta _5 \beta _7 \gamma + 24 \beta _4 \beta _8 \gamma =0, \label{eq7t} \\
	\text{Constants} : &  4 u_0 \beta \beta _1 \beta _7 + 2 \beta _1 \beta _8 + 24 \beta _1 \beta _4 \gamma + 12 \beta _7 ^2 \gamma=0. \label{eq7u}
	\eea 
\end{subequations}
By solving the above system of equations, we get the following set of relations among the parameters: 
\begin{equation}
\begin{aligned}
\beta _1 & = \dfrac{-1875 \beta _3 ^3 \gamma ^3 }{64 u_0 ^6 \beta ^6 } ; \quad \beta _2 = \dfrac{\beta _3 }{2 u_0 \beta} ; \quad \beta _4 = \dfrac{-25 \beta _3 ^3 \gamma
}{16 u_0 ^4 \beta ^4 } ; \quad \beta _5 = \dfrac{-45 \beta _3 ^3 \gamma}{4 u_0 ^3 \beta ^3 } ; \\ \beta _6 &= \dfrac{-17 \beta _3 ^3 \gamma}{4 u_0 ^2 \beta ^2 };  \quad 
\beta _7  = \dfrac{-125 \beta _3 ^3 \gamma ^2 }{32 u_0 ^5 \beta ^5 }; \quad \beta _8 = \dfrac{475 \beta_3^3 \gamma ^2}{16 u_0 ^4 \beta ^4}.  
\end{aligned} \label{eq8}
\end{equation}
From Eqs. \eqref{trans}, \eqref{eq6} and (\ref{eq8}), we obtained the two-rogue wave solution of Benjamin-Ono equation (\ref{eq1.1}) as follows:
\bes\begin{equation}
u(x,t) = u_0 + \dfrac{6 \gamma}{\beta } \left ( \dfrac{F_1}{F_2^2}\right ), \label{eq9}
\end{equation}
where
\begin{eqnarray}
F_1 & =& 4 u_0 \beta ( 4608 t^8 u_0 ^9 x^2 \beta ^9 + 3072 t^{10}u_0 ^{10} \beta ^{10} + 531250 u_0 x^2 \beta \gamma ^4 + 234375 \gamma ^5 \nonumber\\
&&+ 62500 u_0 ^2 \beta ^2 \gamma ^3 ( -2x ^4 + 7 t^2 \gamma )  + 768 t^4 u_0 ^7 x^2 \beta ^7 ( -x^4 + 14 t^2 \gamma ) - 768 t^6 u_0 ^8 \beta ^8 ( -2x^4 + 47 t^2 \gamma )\nonumber\\
&& - 2000 u_0 ^3 x^2 \beta ^3 \gamma ^2 (7 x^4 + 510 t^2 \gamma )  + 400 u_0^4 \beta ^4 \gamma ( 3 x ^8 +210 t^2 x^4 \gamma - 1850 t^4 \gamma^2 )\nonumber\\
&&-96 u_0 ^5 x^2 \beta ^5(x^8+20 t^2 x^4 \gamma -1250 t^4 \gamma ^2)+64 t^2 u_0 ^6 \beta ^6(-9 x^8+90 t^2 x^4 \gamma+2830 t^4 \gamma ^2)),\\
F_2 & =&  96 t^4 u_0 ^5 x^2 \beta ^5 + 64 t^6 u_0 ^6 \beta ^6 - 250 u_0 x^2 \beta \gamma ^2 - 1875 \gamma^3 + 8 u_0 ^3 x^2 \beta ^3 ( x^4 -90 t^2 \gamma )\nonumber\\
&&  - 100 u_0 ^2 \beta ^2 \gamma(x^4 - 19 t^2 \gamma ) + 16 t^2 u_0 ^4 \beta ^4  ( 3x ^4 - 17 t^2 \gamma ).
\end{eqnarray}
\label{eq11}\ees 
\begin{figure}[h] 
	\begin{center} 
		\includegraphics[width=0.85\linewidth]{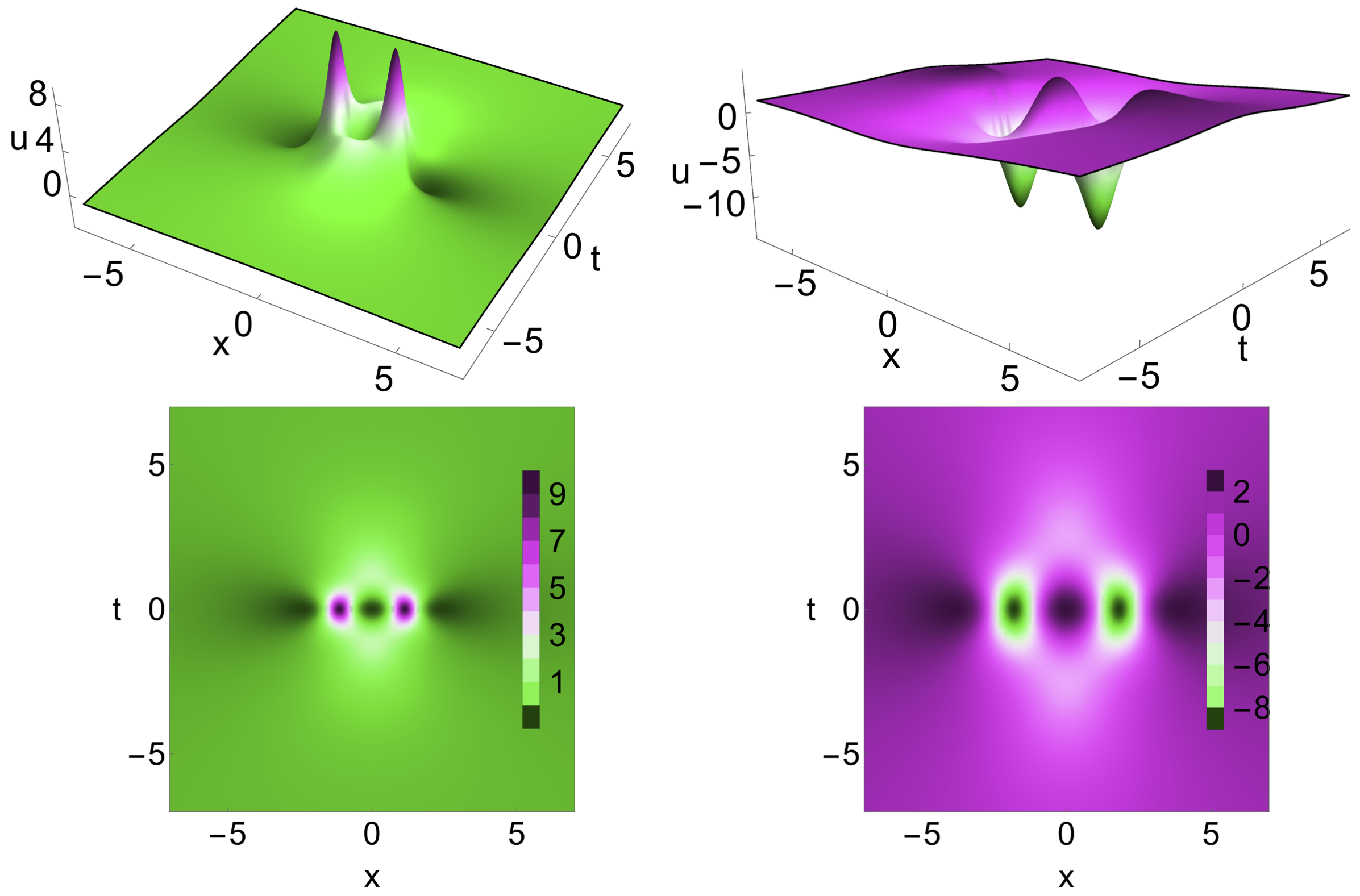}
		\caption{Bright (left panel) and dark (right panel) type second-order rogue waves through solution \eqref{eq11}. The bottom panel shows the corresponding contour plot. The parameter choice for bright rogue wave is $u_0=-1.05$, $\beta=-1.01$, $\beta_3=1.1$ and $\gamma=0.5$, while that of dark rogue wave is $u_0=1.5$, $\beta=0.41$, $\beta_3=1.1$ and $\gamma=-0.75$.}
	\end{center} \label{fig-rogue1c}
\end{figure}
\begin{figure}[h] 
	\begin{center} 
		\includegraphics[width=1.05\linewidth]{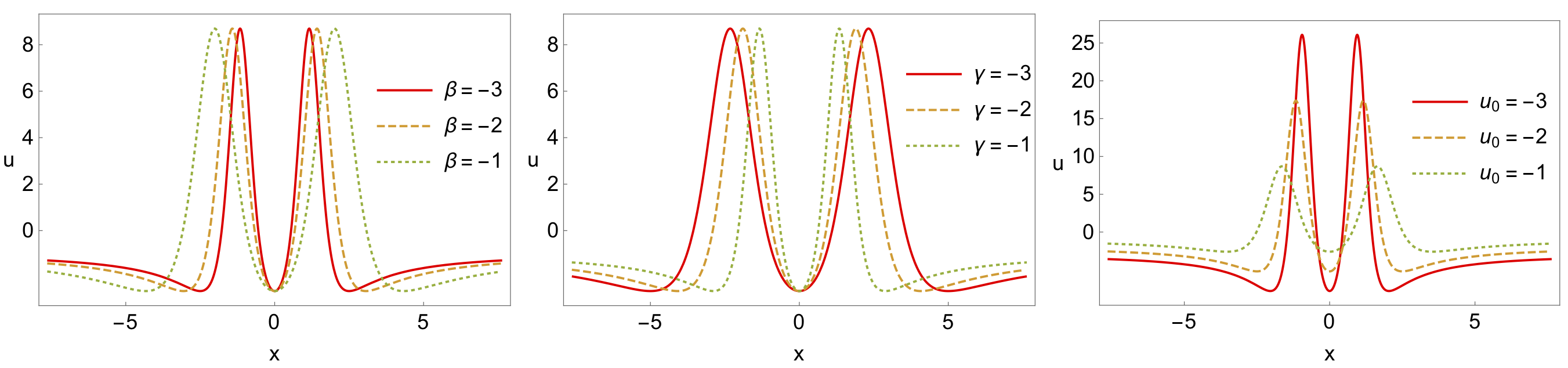}
		\caption{Manipulation of the second-order bright rogue wave \eqref{eq5} by controlling the parameters $\beta$, $\gamma$ and $u_0$.}
	\end{center} \label{fig-rogue2}
\end{figure}

Our categorical analysis on the above second-order rogue wave solution reveals that there exist two types of rogue waves, namely bright and dark rogue waves, as appeared in the case of first-order rogue wave solutions. For completeness, we have shown such a bright and dark type second-order rogue wave structures in Fig. 5 and the respective choices of parameters are given in the caption. Further, the impact of arbitrary parameters $\beta$, $\gamma$, and $u_0$ gives additional freedom to manipulate the amplitude or depth, width, background and tail of the second-order rogue waves too. Such effects of these parameters in the second-order bright rogue wave are demonstrated in Fig. 6. Similarly, these effects can also be observed in the dark rogue wave, which is not given here by considering the length of the article.  

\section{Generalized Breathers}
As given in the introduction, the second part of this work is to investigate the dynamics of generalized breathers of the Benjamin-Ono equation \eqref{eq1.1}, which we carry out in this section. For this purpose, we first obtain the breather solutions of BO model (\ref{eq1.1}) by adopting a generalized three-wave test function suggested in Ref. \cite{ygu}, which is also referred to as the three-wave method. Here the following form of homoclinic breather (two-wave) ansatz \cite{xby} is generalized to obtain the homoclinic breathers as well as travelling wave solutions: 
\begin{equation}
f(x,t) =  m_1 e ^{(p_1 x + k_1 t)} + m_2 \cos (q_1 x+a_1 t)+m_1 e^{-(p_1 x+k_1 t)} , \label{eq31}
\end{equation} 
were $m_1, m_2,  p_1, k_1, q_1, a_1$ are arbitrary constants. The above one is a combination of two waves, a periodic wave $\cos (q_1 x+a_1 t)$ and a hyperbolic wave $\cosh(p_1 x + k_1 t)$, which gives rise to homoclinic breathers. For more generalized breather wave solutions, we consider a combination of three waves as initial test function as given below \cite{ygu}.
\begin{equation}
f(x,t) = b_1 e ^{(px + kt)} + b_2 \cos (qx-at)+b_3 e^{-(px+kt)} + b_4 \cosh (mx +ct), \label{eq33}
\end{equation}
where $b_1,b_2,b_3,b_4,p,k,q,a,m$ and $c$ are unknown parameters to be determined for generalized breathers. Here the choice $b_4=0$ in (\ref{eq33}) correspond to the homoclinic breathers of the two-wave interaction method (\ref{eq31}). Thus, we construct homoclinic breather solutions without loss of generality, which exhibits different nonlinear wave structures ranging from bright/dark solitons, breathers, etc. by following the above three-wave interaction method. It is also clear that the number of arbitrary parameters in the three-wave method is higher than two-wave homoclinic breathers. 

Further, one can also use the following form of the test function, as suggested in Ref. \cite{wta3} for a KdV type system,
\begin{equation}
f(x,t) = b_1 \cosh (\Xi_1)+b_2 \cos (\Xi_2) + b_3 \cosh (\Xi_3), \label{eq3.3}
\end{equation}
where $\Xi_i=k_i(n_i x +r_i y+ s_i t+ \alpha_i), ~i=1,2,3,$ and $n_i , r_i , s_i , \alpha_i$ are the arbitrary parameters. This newly introduced test function is similar to the three-wave solution test function. In a similar way, a generalized N-wave method can also be used to find various solutions of nonlinear integrable/non-integrable equations, which is beyond the scope of the current work and can be investigated separately to look for other types of possible nonlinear wave entities.

\subsection{Breather Solution using Three-Wave method}
Considering the three-wave test function (\ref{eq33}) and collecting the coefficients of $e^{i (px + kt)}$, $\cos (qx-at)$, $\sin (qx-at)$, $\cosh (mx+ct)$, $\sinh (mx +ct), (i=-1,0,1)$, we get the following system of algebraic equations from Eq. (\ref{eq2.1}):
{\small \bes \bea
	(8b_1b_3 k^2 + 16 b_1 b_3 p^2 u_0 \beta + 32 b_1 b_3 p^4 \gamma ) + ( -2a ^2 b_2 ^2 - 4 b_2 ^2 q^2 u_0 \beta + 8 b_2 ^2 q^4 \gamma ) \nonumber\\ + ( 2 b_4 c^2 + 4 b_4 ^2 m^2 u_0 \beta + 8 b_4 ^2 m^4 \gamma )  = 0, \label{eq3.1} \\
	-2a^2 b_2 b_3 + 2 b_2 b_3 k^2 + 4 b_2 b_3 p^2 u_0 \beta - 4 b_2 b_3 q^2 u_0 \beta + 2 b_2 b_3 p^4 \gamma - 12 b_2 b_3 p^2 q^2 \gamma + 2 b_2 b_3 q^4 \gamma  = 0, \label{eq3.2} \\
	-2a^2 b_1 b_2 + 2 b_1 b_2 k^2 + 4 b_1 b_2 p^2 u_0 \beta - 4 b_1 b_2 q^2 u_0 \beta + 2 b_1 b_2 p^4 \gamma - 12 b_1 b_2 p^2 q^2 \gamma + 2 b_1 b_2 q^4 \gamma   =0, \label{eq3.03}\\
	2b_1b_4c^2 + 2 b_1b_4k^2 + 4 b_1 b_4 m^2 u_0 \beta + 4 b_1 b_4 p^2 u_0 \beta ) + (2b_1 b_4 m^4 \gamma + 12 b_1 b_4 m^2 b^2 \gamma + 2 b_1 b_4 b^4 \gamma  =0, \label{eq3.4}\\
	2 b_3 b_4 c^2 + 2 b_3 b_4 k^2 + 4 b_3 b_4 m^2 u_0 \beta + 4 b_3 b_4 u_0 \beta + 2 b_3 b_4 m^4 \gamma + 12 b_3 b_4 m^2 p^2 \gamma + 2 b_3 b_4 p^4 \gamma   = 0, \label{eq3.5} \\
	-2a^2 b_2 b_4 + 2 b_2 b_4 c^2 + 4 b_2 b_4 m^2 u_0 \beta - 4 b_2 b_4 q^2 u_0 \beta + 2 b_2 b_4 m^4 \gamma - 12 b_2 b_4 m^2 q^2 \gamma + 2 b_2 b_4 q^4 \gamma   = 0, \label{eq3.6} \\
	4 ab_1 b_2 k - 8b_1b_2  pq u_0 \beta - 8 b_1 b_2 p^3 q  \gamma + 8 b_1 b_2 pq^3 \gamma   = 0, \label{eq3.7}\\
	-4ab_2 b_3 k + 8 b_2 b_3 pqu_0 \beta + 8 b_2 b_3 p^3 q \gamma - 8b_2 b_3 pq^3 \gamma  =0, \label{eq3.8} \\
	-4b_1 b_4 ck - 8 b_1 b_4 mpu_0 \beta - 8 b_1 b_4 m^3 p \gamma - 8 b_1 b_4 mp^3 \gamma   =0, \label{eq3.9}\\
	4 b_3 b_4 c k+ 8 b_3 b_4 mpu_0 \beta + 8 b_3 b_4 m^3 p \gamma + 8 b_3 b_4 mp^3 \gamma  =0, \label{eq3.10}\\
	4 ab_2 b_4 c - 8 b_2 b_4 mqu_0 \beta - 8 b_2 b_4 m^3 q \gamma + 8 b_2 b_4 mq^3 \gamma  =0. \label{eq3.11}
	\eea \label{eqthree} \ees } 
After solving the above nonlinear algebraic system of equations, different classes of solutions can be obtained for the arbitrary parameters which we discuss one by one in the following part. \\

\noindent 
\textbf{Case 1:} When $ b_1 = b_3 =b_4 =0$ and $b_2 \neq 0$, Eqs. \eqref{eqthree} give $a = \sqrt{2 ( 2q^4 \gamma - q^2 u_0 \beta )}$ with $b_2 $ as an arbitrary free parameter. This results into the following form of $f$:
\begin{equation}
f = b_2 \cos  \left ( qx -  \sqrt{2 ( 2q^4 \gamma - q^2 u_0 \beta ) }t \right ).   \label{eq3.12}
\end{equation}
Thus we get a singular solution of the Benjamin Ono equation (\ref{eq2.1}) as
\begin{equation}
u(x,t) = u_0 - \dfrac{6 \gamma }{\beta } \left (q^2 \sec ^2  (qx - t\sqrt{2 ( 2 q^4 \gamma - q^2 u_0 \beta )  } \right ). \label{eq3.13}
\end{equation}
The above solution always result into unbounded singular form without much advantages or applications. So, here we do not discuss any further details of the obtained solution \eqref{eq3.13}.  \\

\noindent \textbf{Case 2:} When  $b_1 = b_3 =0$ and $b_2, b_4 \neq 0$, the explicit form of $f$ is obtained from Eqs. \eqref{eqthree} as
\bes \begin{align}
f= b_2 \cos \left [  \left ( \frac{\sqrt{(u_0 \beta+2m^2\gamma)}}{\sqrt{2 \gamma}}\right)\left(\sqrt{3}x-2m \sqrt{\gamma}t\right) \right ]+ \frac{2 \sqrt{ -2b_2^2 (u_0 \beta + 2 m^2 \gamma }}{\sqrt{u_0 \beta + 8 m^2 \gamma}}\nonumber\\
\times\cosh \left [ mx - \sqrt{\dfrac{3}{4\gamma}} \left ( {u_0 \beta + 4 m^2 \gamma} \right)t \right ],\label{eq2.16}
\end{align}
where 
$b_4 = 2 \sqrt{ \dfrac{ -2b_2^2 (u_0 \beta + 2 m^2 \gamma)}{u_0 \beta + 8 m^2 \gamma}}$, 
$q  = \sqrt{\dfrac{3 (u_0 \beta +  2 m^2 \gamma)  }{2 \gamma}}$,
$c  = - \sqrt{\dfrac{3}{4\gamma}} {(u_0 \beta + 4 m^2 \gamma )}$, and
$a  =  \sqrt{2m^2 ( u_0 \beta + 2 m^2 \gamma )}$ while the other parameters ($u_0,~\beta,~\gamma,~b_2$ and $m$) are arbitrary. From the above $f$ and Eqn. \eqref{trans}, we can obtain the exact solution as below.
\bea
u(x,t) = u_0 + \dfrac{6 \gamma}{\beta} ( ln \, f)_{xx}.
\eea \label{eq2.16aa} \ees
\begin{figure}[h] 
	\begin{center} 
		\includegraphics[width=1.01\linewidth]{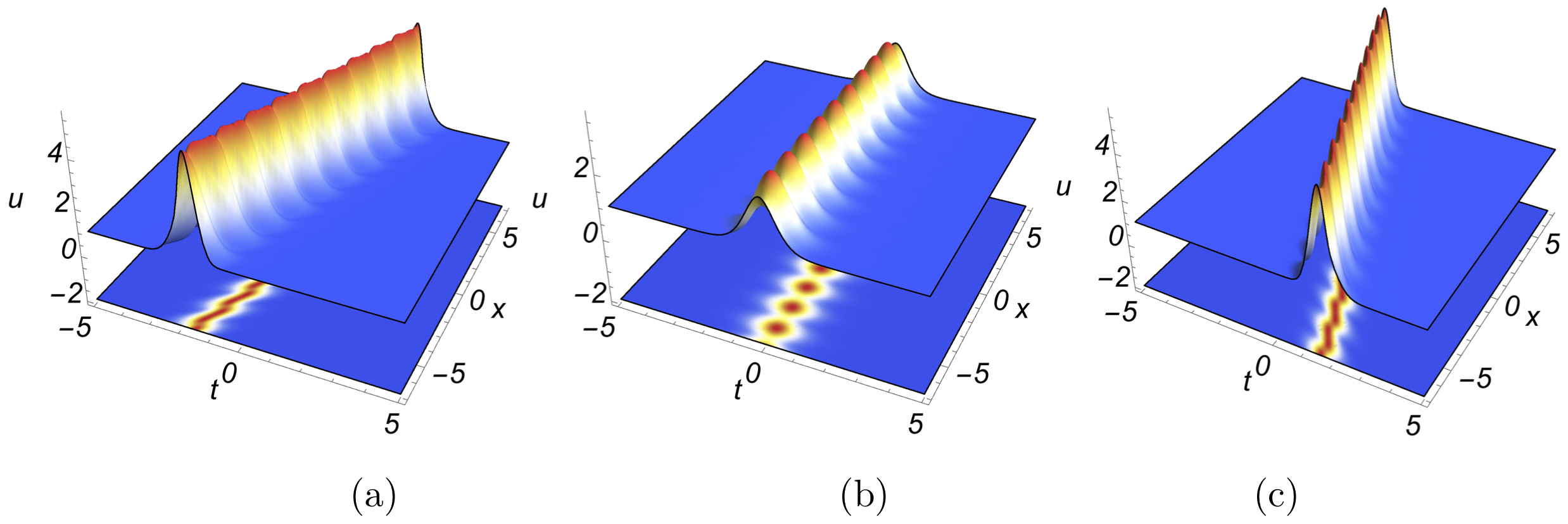}
		\caption{Breathing soliton with periodical oscillation in amplitude and space for (a) $m = 0.75$, (b) $m = 0.0$ and (c) $m = -0.75$. Other arbitrary parameters are fixed as $u_0 = 1$, $\gamma = 0.5$, $\beta = 1.50$, and $b_2 = 1.0$.}
	\end{center} \label{figbreath1}
\end{figure}

This clearly shows that the contribution from both `cos' and `cosh' parts giving rise to the breathing nature of soliton, which may be of either bright (zero background) or anti-dark (non-zero background)  solitons for an appropriate choice. Further, the breathing (period of) oscillations can be controlled in addition to the manipulation of breather velocity (direction of propagation), amplitude and width by tuning the available arbitrary parameters. To be specific, the $u_0$ parameter controls the background energy and amplitude of the breather, while $\gamma$ influences the amplitude and velocity. However, the parameter $\beta$ varies the width of the soliton breather in addition to its velocity changes. For illustrative purposes, we have given a bright soliton breather on a constant non-zero energy/amplitude (anti-dark soliton breathers) in Fig. 7 travelling with positive, zero, and negative velocity having different amplitudes. 

Another striking feature in the above obtained solution (\ref{eq2.16aa}), which is nothing but the interaction of two stable solitons and it occurs when there exist the background amplitude/intensity. This includes the collision between (i) two single-hump (anti-dark) solitons, (ii) single-well (dark) and single-hump (anti-dark) solitons, (iii) rational type and dark solitons, and (iv) bound state formation among two anti-dark solitons by choosing the appropriate choices of parameters as shown in Fig. 8 for elucidation. Here we have depicted the elastic (amplitude or shape preserving) collision of two non-equal amplitude anti-dark solitons (Fig. 8a), while such elastic collision can also be possible among two equal amplitude solitons. We can easily witness that though the solitons are interacting elastically, they undergo a phase-shift after collision (clearly visible in contour plots). Soliton bound states are nothing but two solitons travelling with same velocity (velocity resonance) exhibit periodic attraction and repulsion of their central position during propagation. This can also happen in higher-dimensional solitons as well as among solitons undergoing an inelastic collisions . For more detailed investigation on soliton collisions and bound states please refer \cite{soli-int1,soli-int2,soli-int3,bound1,bound2,bound3} and references therein.\\  
\begin{figure}[h] 
	\begin{center} 
		\includegraphics[width=1.01\linewidth]{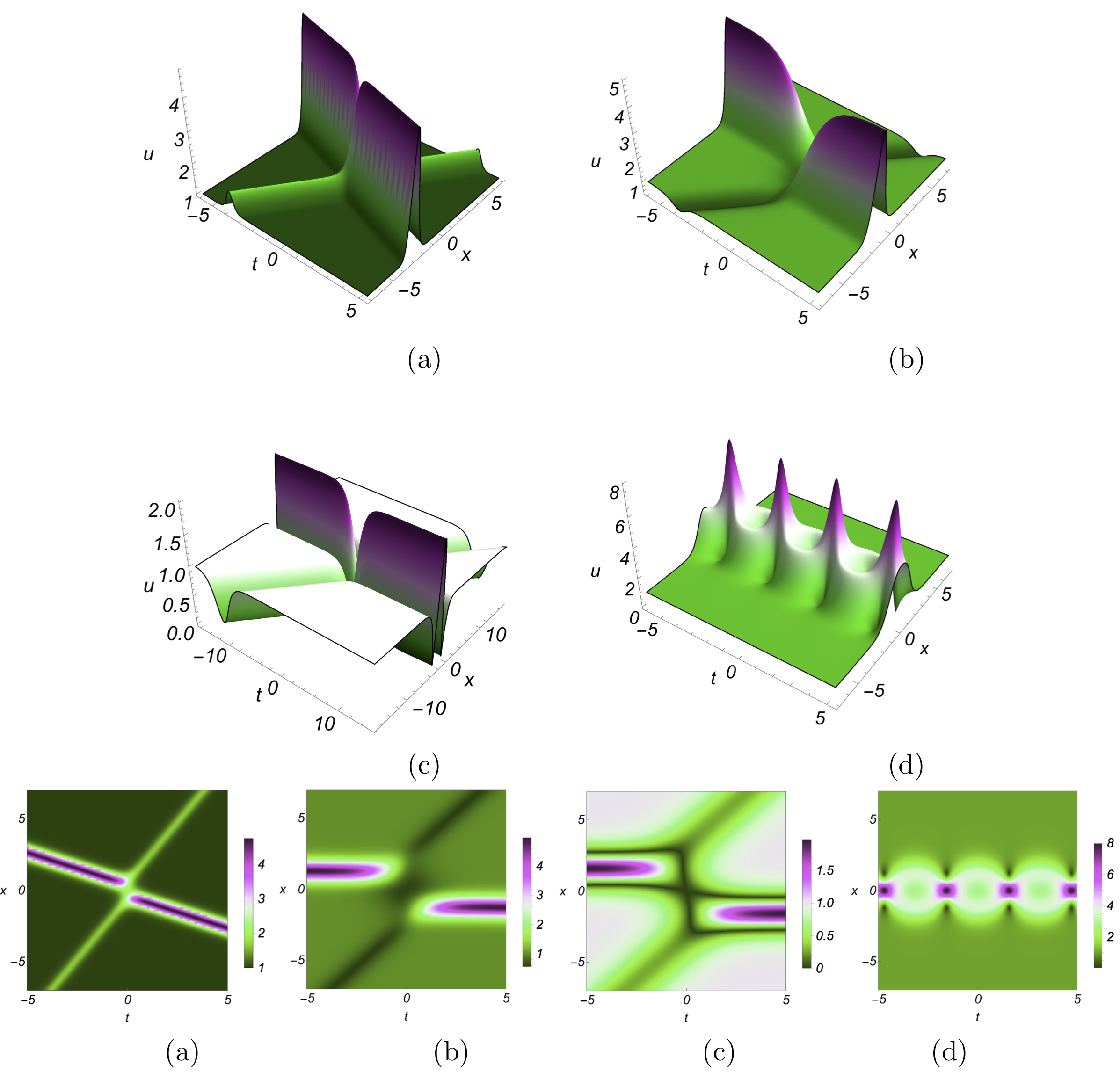}
		\caption{Elastic interaction between (a) among two single-hump (anti-dark)  solitons, (b) anti-dark and dark solitons, (c) dark and rational solitons, and (d) bound state of two anti-dark solitons  through (\ref{eq2.16aa}). Bottom panel shows the contour plot of these interacting solitons. The parameter choices are (a) $u_0 = 1.0$, $\beta = -2.5$, $\gamma = 0.5$, $b_2=1.0$, $p=-1.0$, \& $m=0.5$, (b) $u_0 = 1.0$, $\beta = -1.15$, $\gamma = 0.5$, $b_2=1.0$, $p=-1.0$, \& $m=0.5$, (c) $u_0 = 1.0$, $\beta = -0.95$, $\gamma = 0.5$, $b_2=1.0$, $p=-1.0$, \& $m=0.5$, and (d) $u_0 = 1.0$, $\beta = 1.0$, $\gamma = -0.75$, $b_2=1.0$, $p=-1.0$, \& $m=0.0$.} 
	\end{center} \label{figure-breath}
\end{figure}

\noindent 
\textbf{Case 3:} For the choice $b_2 = b_4 =0$ and $b_1, b_3 \neq 0$, Eqs. \eqref{eqthree} reduces to an explicit form of $f$ as
\begin{equation}
f = b_1 e ^{(px - \sqrt{2 (-p^2 u_0 \beta - 2 p^4 \gamma)}t)}    + b_3 e ^{-(px - \sqrt{2 (-p^2 u_0 \beta - 2 p^4 \gamma } t )}, \label{eq3.15}
\end{equation}
along with $k = -\sqrt{ -2 p^2 (u_0 \beta +2 p^2 \gamma )}$. Further, by setting $b_1 =b_3 >0$, the function $f(x,t)$ becomes 
\bes \begin{equation}
f(x,t) =2 b_1 \cosh \left( px - \sqrt{ -2 p^2 (u_0 \beta +2 p^2 \gamma  } t \right) .\label{eq3.16}
\end{equation}
Hence the solution of BO equation (\ref{eq1.1}) is obtained as the following stable soliton:\\
\begin{equation}
u(x,t)=u_0 + \dfrac{6 \gamma }{\beta }p^2  \sech ^2\left( px - t\sqrt{ -2 p^2 (u_0 \beta +2 p^2 \gamma  } )\right).\label{eq3.17}
\end{equation}\ees 
On the other hand, when $b_1=d_1, b_3=-d_1$ with $d_1>0$ the function $f(x,t)$ reduces to 
\bes \begin{equation}
f(x,t) =2 d_1 \sinh \left( px - \sqrt{ -2 p^2 (u_0 \beta +2 p^2 \gamma  } t \right). \label{eq3.18}
\end{equation}
Now, the corresponding solution of BO equation (\ref{eq1.1}) can be derived as a hyperbolic solution given below.
\begin{equation}
u(x,t)=u_0 - \dfrac{6 \gamma }{\beta }p^2 \mbox{cosech} ^2\left( px - t\sqrt{ -2 p^2 (u_0 \beta +2 p^2 \gamma} )\right).\label{eq3.19}
\end{equation}\ees 

From the above solutions, namely Eqs. \eqref{eq3.17} and \eqref{eq3.19}, one can understand that they correspond to bright/dark soliton and unbounded structures, respectively. Though the singular solutions are of no further interest, solitons found multifaceted applications due to their stable propagation with various localized profiles of salient features. In the present case, the solution \eqref{eq3.17} is further divided into two categories: (i) bright solitons (localized stable waves appearing on a zero-background $u_0=0$) and (ii) dark/gray/anti-dark solitons (stable localized waves arising on a non-zero background $u_0\neq 0$). In bright soliton case ($u_0=0$), an additional restriction on the parameters ($\gamma <0$) also arise, which overcomes/removes the singular structures. Such bright solitons travel with velocity $\sqrt{ -4 p^4 \gamma}$ and possess an amplitude of $\dfrac{6 \gamma }{\beta }p^2$. The second case ($u_0\neq 0$), we obtain various profile structures ranging from anti-dark, dark, gray and rational type solitons for different choices of other parameters provided they satisfy $u_0 \beta +2 p^2 \gamma<0$. These nonlinear wave structures admit an amplitude $u_0+\dfrac{6 \gamma }{\beta }p^2$ and propagates with velocity $\sqrt{ -2 p^2 (u_0 \beta +2 p^2 \gamma)  }$. Among these, the anti-dark soliton resembles with standard bright solitons, but it appear on the constant background and with different velocity. On the other hand, the dark solitons can also be divided into two sub-categories called dark and gray solitons, where the former localized waves have zero lowest amplitude (depth of the well structure), while in the latter they admit non-zero well depth. Additionally, we have identified bright and dark type rational solitons by controlling the arbitrary parameter $p$. Here the dark rational solitons can be viewed as a double-well (W-shaped) dark solitons, while the bright rational soliton is nothing but a localized hump/peak with side-band minima on either side of localization. By preserving these relations and tuning the available parameters, one can manipulate their width and amplitude along with propagation direction appropriately. For completeness and better understanding, we have shown such bright, anti-dark, dark, and gray solitons in addition to bright and dark rational solitons in Fig. 9. \\ 
\begin{figure}[h] 
	\begin{center} 
		\includegraphics[width=1.01\linewidth]{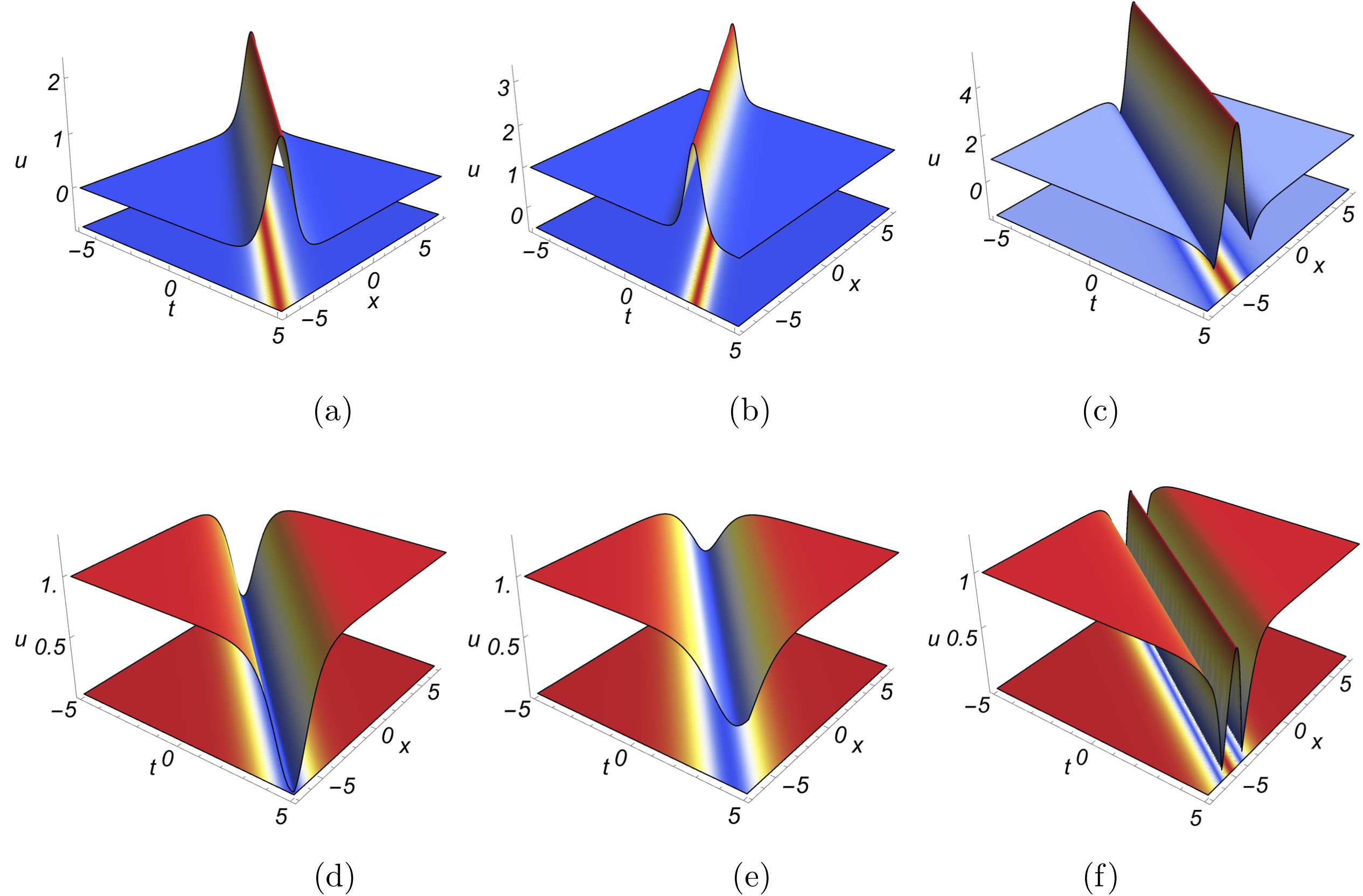}
		\caption{Propagation of (a) bright soliton, (b) anti-dark soliton (bright soliton on a constant background $u_0\neq 0$), (c) rational soliton, (d) single-well dark soliton, (e) single-well gray solitons, and W-shaped (double well) dark soliton of BO equation through Eq. \eqref{eq3.17}. The parameters are chosen as (a) $u_0 = 0$, $\gamma = -0.5$, $\beta = -1.5$, $d_1 = 1.0$, \&  $p = -1.0$, (b) $u_0 = 1.0$, $\gamma = -0.5$, $\beta = -1.5$, $d_1 = 1.0$, \&  $p = -1.0$,  (c) $u_0 = 1.0$, $\gamma = -0.5$, $\beta = 0.5$, $d_1 = 1.5$, \&  $p = -1.0$, (d) $u_0 = 1.0$, $\gamma = 0.5$, $\beta = -1.5$, $d_1 = 1.0$, \&  $p = -0.7$, (e) $u_0 = 1.0$, $\gamma = 0.5$, $\beta = -1.5$, $d_1 = 1.0$, \&  $p = -0.5$, and (f) $u_0 = 1.0$, $\gamma = 0.5$, $\beta = -1.5$, $d_1 = 1.0$, \&  $p = -1.0$.}
	\end{center} \label{sol3-soli3}
\end{figure}

\noindent 
\textbf{Case 4:} When  $b_1 = 0$ and $b_2, b_3, b_4 \neq 0$, we can obtain the explicit form of $f$ from the set of equations \eqref{eqthree}  as below.
\bes \bea
f(x,t) = b_2 \cos (qx-at) + b_3 e^{-(px + kt)} + b_4 \cosh (mx+ct),\label{eq3.17a}
\eea
where the parameters $m,~p,~k,~a$ and $q$ take the following form:
\bea
&&m = \dfrac{1}{2 \sqrt{\gamma} } \sqrt{-u_0 \beta - \Delta  } ,\quad 
p  = \sqrt{\dfrac{3}{4\gamma}} \sqrt{-u_0 \beta + \Delta},  \\ 
&&k = \frac{- \sqrt{3}\Delta}{4 c \gamma} \sqrt{-(u_0 \beta + \Delta)} \sqrt{-(u_0 \beta - \Delta)}, \qquad \quad\\
&&a = \dfrac{1}{4 c \gamma}{(u_0 \beta - \Delta) \sqrt{-(u_0 \beta + \Delta)^2}},  \qquad\\
&& q = \dfrac{1}{2 \sqrt{\gamma}} \sqrt{u_0 \beta + \Delta}, \quad \mbox{with} \quad \Delta =\sqrt{u_0^2 \beta ^2 - 4 c^2 \gamma}.
\eea \label{case4sol}\ees
\begin{figure}[h]
	\begin{center} 
		\includegraphics[width=1.01\linewidth]{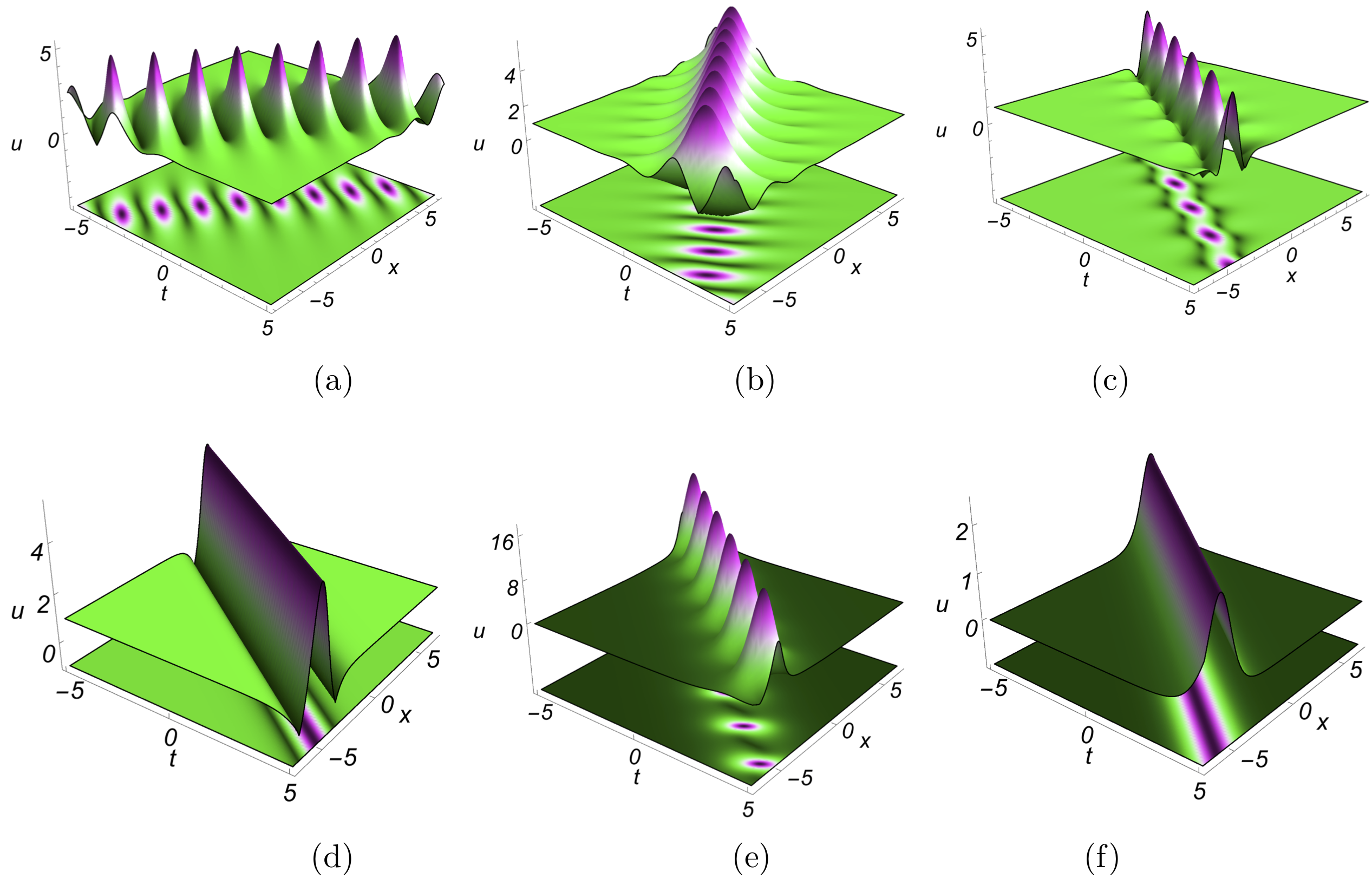}
		\caption{Propagation of (a-d) soliton breathers appearing on a constant background, and (e) breathing and (f) stable bright solitons on zero-background on through \eqref{case4sol}. The parameter choices are (a) $u_0 = 1$, $\beta = -1.5$, $\gamma = -0.5$, $b_2=1.0$, $b_3=0.5$, $b_4=1.5$, \& $c=-1.0$, (b) $u_0 = 1$, $\beta = -1.5$, $\gamma = -0.5$, $b_2=-1.0$, $b_3=-0.5$, $b_4=1.5$, \& $c=1.0$, (c) $u_0 = 1$, $\beta = 1.5$, $\gamma = -0.75$, $b_2=1.0$, $b_3=0.5$, $b_4=1.5$, \& $c=1.0$, (d) $u_0 = 1$, $\beta = 0.5$, $\gamma = -0.5$, $b_2=0.5$, $b_3=0.0$, $b_4=0.5$, \& $c=1.0$, (e) $u_0 = 0.0$, $\beta = 1.0$, $\gamma = -0.5$, $b_2=0.5$, $b_3=-0.5$, $b_4=-1.5$, \& $c=1.0$, and (f) $u_0 = 0.0$, $\beta = 1.0$, $\gamma = -0.5$, $b_2=0.5$, $b_3=0.0$, $b_4=-1.5$, \& $c=1.0$.} 
	\end{center} \label{figure1}
\end{figure}

In the above solution \eqref{case4sol}, $u_0$, $\beta$, $\gamma$, $c$, $b_2$, $b_3$, and $b_4$ are the seven arbitrary parameters through which we can manipulate the resultant nonlinear wave pattern of soliton breather. Compared to the previous solutions, the above solution reveals exciting patterns of soliton breather explaining various phenomena. It starts from the breathing of solitons with periodic oscillations in its amplitude as well as stable rational soliton on non-zero background $u_0\neq 0$ and stable/standard bright soliton without any background ($u_0= 0$). Further, we have also identified solitons with single or double hump/well structure undergoing fusion and fission processes in addition to their bending characteristics. Such type of soliton breather is shown in Fig. 10, while the bending, fission, and fusion nature of solitons are demonstrated in Figs. 11--13 with an appropriate choice of arbitrary parameters. Here one can also control the amplitude, period of oscillations/breathing, width, and velocity of solitons/breathers by tuning the arbitrary parameters. 
\begin{figure}[h] 
	\begin{center} 
		\includegraphics[width=0.75\linewidth]{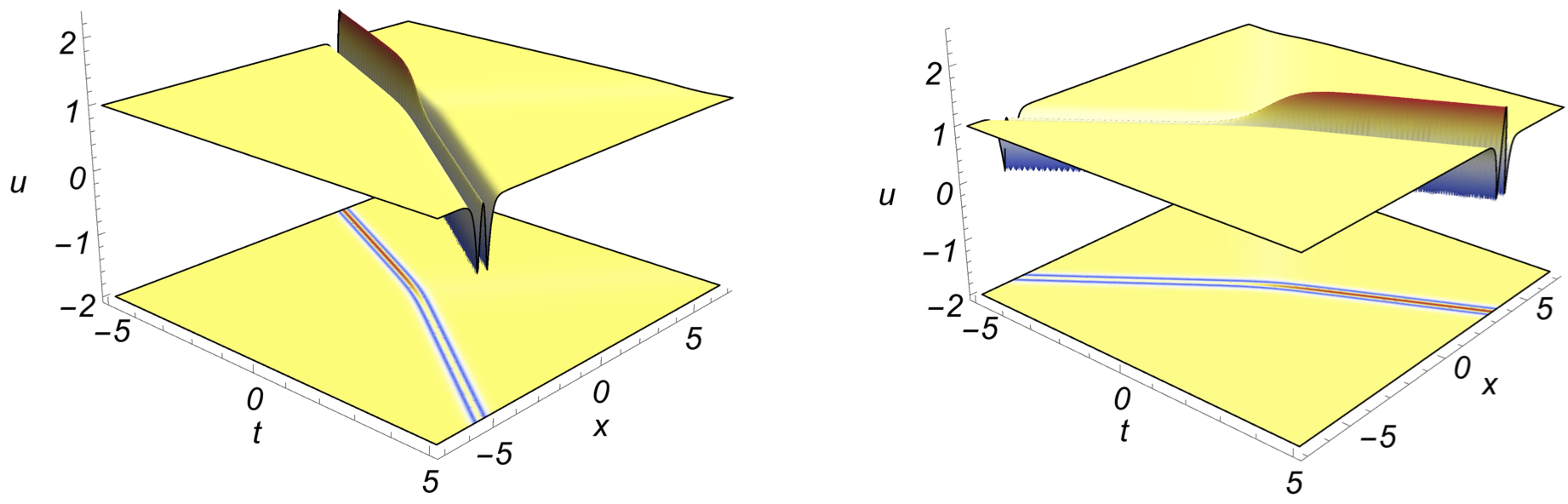}
		\caption{Bending of W-shaped dark soliton with decrease/increase in the intensity due to increased/decreased width of the soliton given by Eqn. \eqref{case4sol}  for $c =- 0.5$ (left panel) and $c=0.5$ (right panel) with other arbitrary parameters as $u_0 = 1$, $\beta = -1.5$, $\gamma = 0.25$, $b_2=1.0$, $b_3=0.5$, and $b_4=1.5$.}
	\end{center} \label{figure1-bend}
\end{figure} \begin{figure}[h] \vspace{-0.52cm}
	\begin{center} 
		\includegraphics[width=0.75\linewidth]{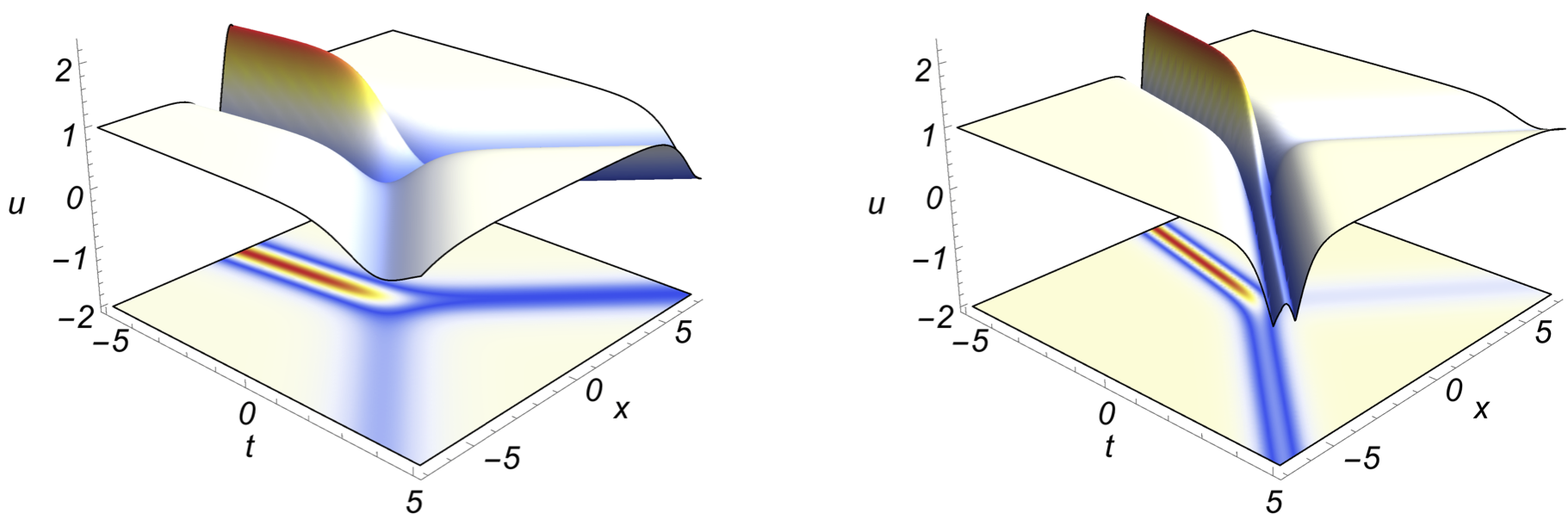}
		\caption{Fission of a W-shaped dark soliton into two single well solitons (for $u_0 = 1$, $\beta = -1.5$, $\gamma = 0.5$, $b_2=1.0$, $b_3=0.5$, $b_4=1.5$, and $c=1.0$) and a single well soliton and double well gray solitons (for $u_0 = 1$, $\beta = -1.5$, $\gamma = 0.25$, $b_2=1.0$, $b_3=0.5$, $b_4=1.5$, and $c=-1.0$) through Eqn. \eqref{case4sol}.}
	\end{center} \label{figure1-fission}
\end{figure}\begin{figure}[h] \vspace{-0.52cm}
	\begin{center}  
		\includegraphics[width=0.75\linewidth]{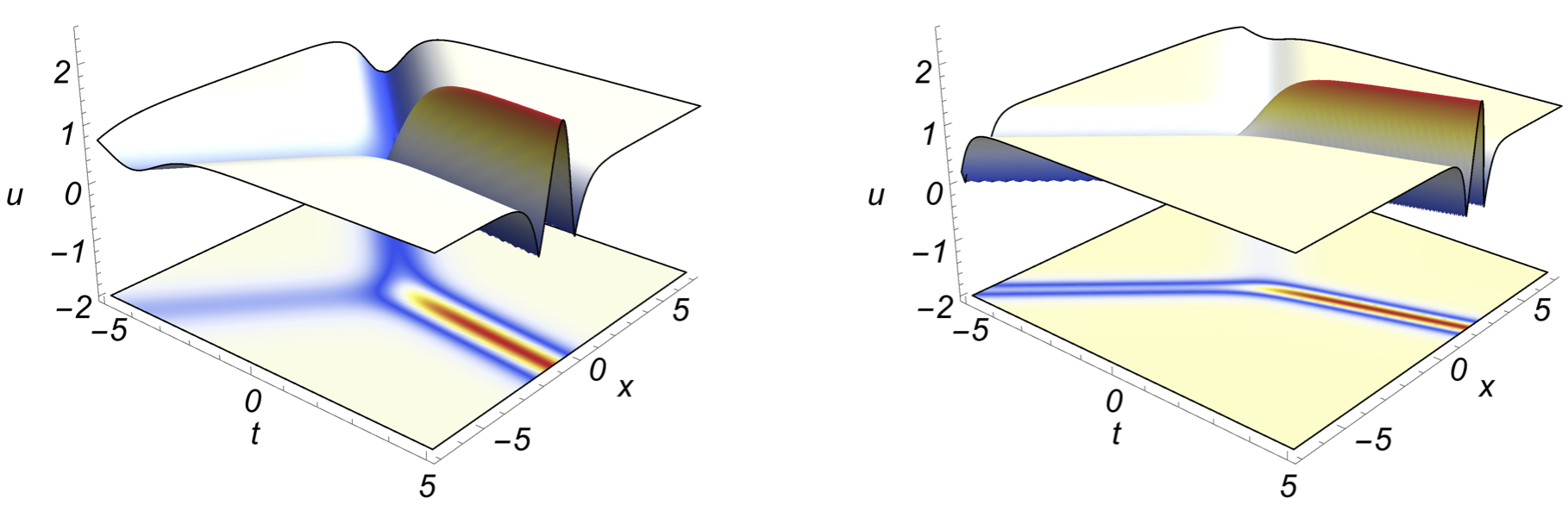}
		\caption{Fusion of two single well solitons (for $u_0 = 1$, $\beta = -1.5$, $\gamma = 0.5$, $b_2=1.0$, $b_3=0.5$, $b_4=1.5$, and $c=1.0$) and a single well soliton with double well soliton (for $u_0 = 1$, $\beta = -1.5$, $\gamma = 0.25$, $b_2=1.0$, $b_3=0.5$, $b_4=1.5$, and $c=1.0$) to form amplified W-shaped dark soliton through Eqn. \eqref{case4sol}.}
	\end{center} \label{figure1-fusion}
\end{figure}

As a future study, the present three-wave interaction (test function) method can be further extended/generalized to arbitrary $N$-wave interaction, which shall reveal several interesting features/dynamics including the interaction among breathers as well as the collision of solitons with breathers. Considering the computational complexity and length of the manuscript, we have not studied here, but it will be worth to explore. 

\section{Conclusion} 
We have considered the Benjamin-Ono equation and constructed various localized wave solutions starting from the rogue waves to breathers and solitons by employing polynomial test functions and  three-wave method with the aid of bilinear form. Through the obtained solutions, we are able to control and manipulate the constructed localized waves with the available arbitrary parameters to realize their multifaceted nature such as tailoring of their amplitude, width, velocity, and tail/valley of both bright and dark type first as well as second-order rogue waves, developing bright/anti-dark, dark/gray, and rational solitons, the interaction of dark, anti-dark, and rational solitons, exploring the formation of bound states, fusion, fission and bending properties of solitons with clear graphical demonstrations. The reported results will be helpful for a complete understanding of the dynamics of the considered Benjamin-Ono model, and further, the analysis can be extended to other related nonlinear models. 

\setstretch{1.150}
\section*{Acknowledgments}
SS is grateful to Ministry of Human Resource Development (MHRD), Govt. of India, and National Institute of Technology, Tiruchirappalli, India, for financial support through institute fellowship. The work of KS was supported by Department of Science nd Technology - Science and Engineering Research Board (DST-SERB), Govt. of India, National Post-Doctoral Fellowship (File No. PDF/2016/000547). Authors thank the anonymous reviewer for providing fruitful comments for the betterment of the manuscript.\\

\noindent{\bf Declarations}\\
\noindent{\bf Conflict of Interest}: The authors declare that there is no conflict of interests regarding the research effort and the publication of this paper.\\~\\
\noindent{\bf CRediT Author Statement}: Sudhir Singh: Conceptualization; Methodology; Writing - original draft, review and editing. K. Sakkaravarthi: Formal analysis; Investigation; Visualization; Writing - original draft, review and editing. K. Murugesan: Funding acquisition; Resources; Supervision; Project administration.  R. Sakthivel: Resources; Supervision; Project administration.


\end{document}